\documentstyle[mprocl,epsfig]{article}

\def\Journal#1#2#3#4{{#1} {\bf #2}, #3 (#4)}

\def\APJ{{\em ApJ.}}
\def\CQG{{\em Class. Quant. Grav.}}
\def\NAT{{\em Nature}}

\def\PRD{{\em Phys. Rev.} D}

\def\be{\begin{equation}}
\def\ee{\end{equation}}
\def\bea{\begin{eqnarray}}
\def\eea{\end{eqnarray}}

\begin{document}

\title{Nonlinear wave equations for
numerical relativity: towards 
the computation of gravitational wave forms of
black hole binaries}

\author{ Maurice H.P.M. van Putten}

\address{Massachusetts Institute of Technology, Cambridge, MA 02139}

\maketitle\abstracts{
The prime candidate of LIGO/VIRGO sources of
gravitational waves is the spiral in of
black holes and neutron stars in compact
binaries. While the early stages of the evolution
of compact binaries is computable from post-Newtonian
calculations, prediction of their late stages requires
large scale numerical simulation. A fully covariant
and strictly hyperbolic formulation for numerical
relativity is described, and illustrated in a
one-dimensional computation of a
Gowdy-wave on the three-torus. This formulation
allows foliations in full generality, 
in particular it poses no restriction on the
lapse function.}
  
\section{Introduction}

The advanced stages of construction of
LIGO and VIRGO has stimulated 
some new developments in hyperbolic
formulations for numerical relativity\cite{mvp1,Abr}.
These developments describe nonlinear
gravitational wave motion in the
late stages of spiral in of compact
binaries of black holes and neutron stars.
Binary black hole and neutron star systems remain today 
the prime
candidates for astrophysical sources
of gravitational radiation,
and statistical calculations
indicate their event rates to be
more than a few per year at the level
of the sensitivity of the initial LIGO
\cite{ph,na}.
Other known 
burst sources with potentially
detectable levels of gravitational
radiation are the collapse of white
dwarfs into neutron stars\cite{usov}, and
secularly unstable single neutron
stars\cite{lai}. 

\section{$\dot{f}(f)-$diagram for
burst sources}

The time evolution of the above mentioned three
classes of burst sources of gravitational radiation
can be illustrated (at the present
level of our understanding) in an
$\dot{f}(f)-$diagram (Figure 1). Recall
that compact binaries show 
an initial
spiral in set by the Newtonian
relationship
\begin{eqnarray}
\dot{f}\sim f^{\frac{11}{3}},
\end{eqnarray}
where $\dot{f}=\mbox{d}f/\mbox{d}t$.
The frequency $f$ refers to the 
base
frequency of the gravitational 
wave-form, initially of the
``chirp," 
and finally of the quasi-normal mode 
ring down. During the intermediate
merger phase, 
$f$ and $\dot{f}$ either follow
by interpolation, or from 
$f=\frac{1}{\pi}\dot{\phi}$ with
$\phi$ 
the orbital phase. Numerical relativity
on black hole binaries specifically
targets the merger phase, where post-Newtonian
calculations break down. The present challenges 
are found to be the computation of stable, multiple
orbits and the treatment of horizon boundary
conditions.
\begin{center}
  \epsfig{file=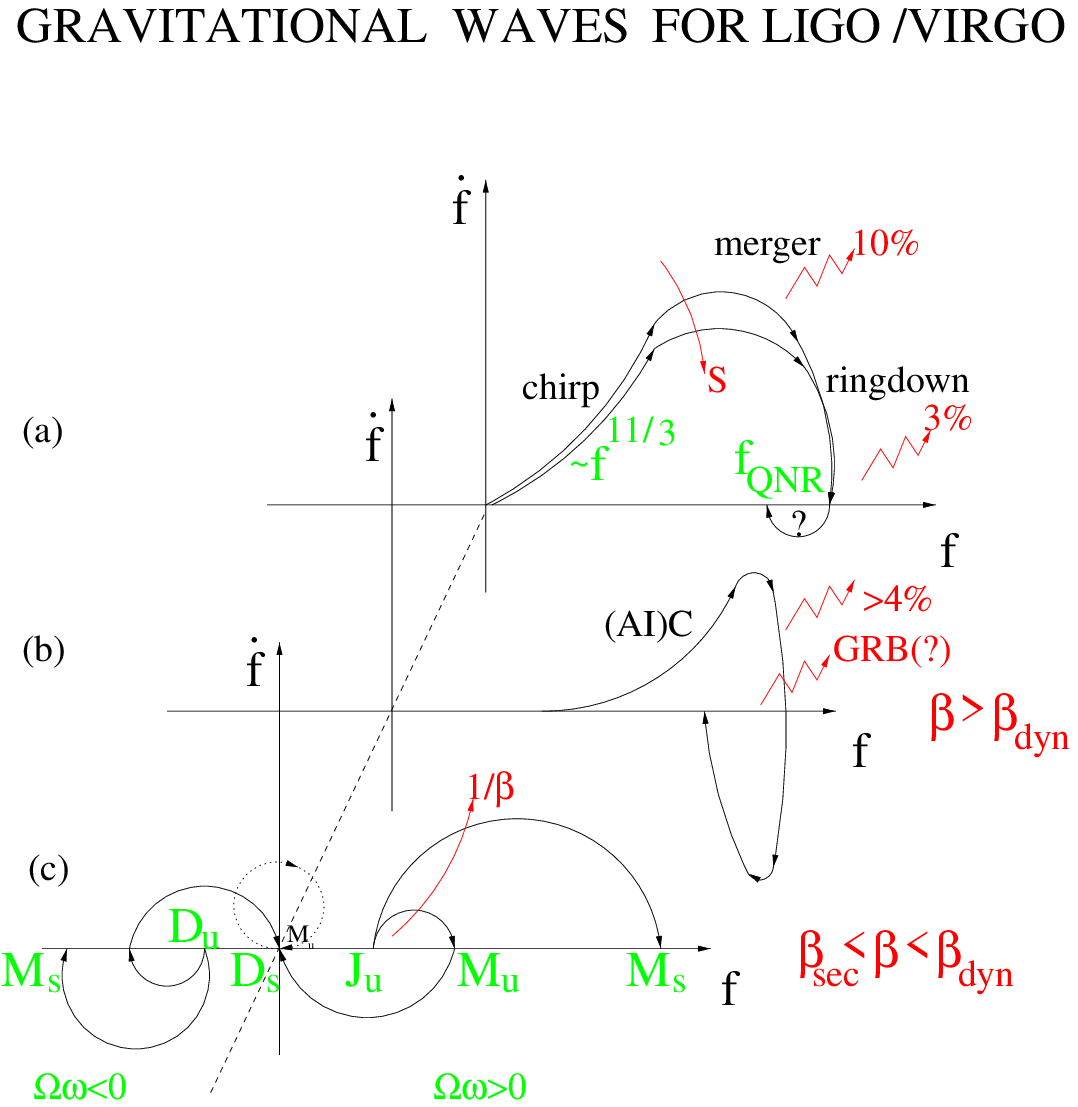,width=105mm,height=85mm}
\end{center}
{\bf FIGURE 1.}
{\small 
Three cartoons of burst sources of gravitational
waves in an $\dot{f}(f)$-diagram.
The {\bf top} diagram shows
the three phases of binary
coalescence: the chirp (with the Newtonian result 
$\dot{f}(f)\sim
f^{11/3}$), the merger phase, and the ring down phase with
quasi-normal mode frequency $f_{QNR}\approx865 \frac{37M}{M}$ Hz
(see, $e.g.$, Flanagan 
\& Hughes\cite{fla}).
The merger phase of two black holes and two
neutron stars is expected to be 
distinct, but in as yet unknown ways. Sketched is the
expected trend with increasing spin $S$ of the binary
objects. 
Order of magnitude 
estimates\cite{fla,tho,sch}
of the radiated energy 
in the merger phase is also indicated.
(A possible break up and disk 
formation in neutron star mergers
has been by 
Zhuge\cite{zhu}, and could also
be included here.)
The {\bf middle} diagram sketches the expected 
gravitational wave form resulting from a
dynamical instability in an (accretion induced)
collapse scenario, such as the collapse
of a white dwarf 
to a neutron star. 
The arrow indicates the
direction of time in this trajectory. The first
stage shows a rapid sweep of increasing 
frequency,
followed by a phase
of rapid damping due to 
gravitational radiation. This scenario
assumes the ratio $\beta$ of
kinetic energy to gravitational energy to exceed
the value which triggers a 
dynamical instability ($\beta>0.27$).
It has been suggested as a model for GRBs by
Usov\cite{usov}. The {\bf lower} diagram
sketches a translation of 
results on the secular instability
of single neutron stars, $\beta_{\mbox{sec}}(=0.1375)
<\beta<\beta_{\mbox{dyn}}(=0.27)$, 
of Lai \& Shapiro\cite{lai}
into the $\dot{f}(f)-$diagram.
The secular instability describes gravitational
radiation driven
transitions from unstable Jacobi 
($J_u$) ellipsoids to either stable Mclaurin
spheroids ($M_s$) or stable Dedekind ellipsoids
($D_s$). The Jacobi ellipsoids are 
rotation $(\Omega)$ dominated (rel. vorticity),
and the Dedekind ellipsoids are vorticity
($\omega$) dominated (rel. rotation) stars. 
The trajectories are extended to the left,
where the shape is in counter rotation
relative to the fluid's vorticity. 
Note that the end point of the
trajectory in the middle diagram
connects to $J_u$ of the lower diagram.}

\section{Nonlinear wave equations for relativity}

Fully covariant, nonlinear wave equations have recently been
obtained general relativity in the tetrad approach. They
are Yang-Mills type equations for the Utiyama-connections
$\omega_{a\mu\nu}=(e_\mu)_c\nabla_a(e_\nu)^c$ (for a tetrad
field $\{(e_\mu)_a\}_{\mu=1}^4$) in the Lorentz gauge
$\nabla^a\omega_{a\mu\nu}$. In vacuo, they reduce to
\begin{eqnarray}
\hat{\Box}
\omega_{a\mu\nu}-
[\omega^c,\nabla_a\omega_{c}]_{\mu\nu}=0,
\end{eqnarray}
were $\hat{\Box}=
\hat{\nabla}_c\hat{\nabla}^c$ is the
$SO(3,1)$ gauged wave-operator.
The full set of equations further involves the equations of
structure, which can be integrated in the form of
\begin{eqnarray}
(e_\mu)_a(\tau)=\Lambda_\mu^{\hskip.1in\nu}(\tau,\tau_0)(e_\mu)_c(\tau_0)
+\int_{\tau_0}^\tau \Lambda_\mu^{\hskip.1in\nu}(\tau,s)\hat{\partial}_b
N_\nu(s)\mbox{d}s,
\end{eqnarray}
where the spatial coordinates have been suppressed. Here, the
\begin{eqnarray}
N_\mu=(e_\mu)_\tau
\end{eqnarray}
are the {\em tetrad lapse functions}, which are algebraically
equivalent to the Hamiltonian lapse and shift functions. Strict
hyperbolicity is maintained with 
arbitrary $N_\mu$, provided they define space-like foliations.
There is full generality in this regard. 
By comparison, the otherwise closely related 
hyperbolic formulation 
of Abrahams $et$ 
$al.$\cite{Abr} shows
some restrictions 
on the Hamiltonian lapse function for 
maintaining strict hyperbolicity.
(This difference was left unnoticed in the
comparison by Finn\cite{Finn}).

A numerical implementation is 
given in van Putten,\cite{mvp2}
and applied to a nonlinear Gowdy-wave in a Gowdy $T^3$-cosmology.
Stable and accurate results are obtained with
proper second-order convergence by comparison with analytical
solutions.
No indication of any  
``coordinate shocks" 
as reported by Alcubierre \& Masso
\cite{Alc} 
have been observed in these
numerical experiments.

\end{document}